\documentclass[twocolumn,superscriptaddress,prb,aps]{revtex4}
\usepackage{graphicx}
\usepackage{amsmath}

\begin{document}
\title{Vibrational assignments and line shapes in inelastic
  tunnelling spectroscopy: H on Cu(100)}
\affiliation{
Institute of Physics, Tampere University of Technology,
33720 Tampere, Finland}
\affiliation{
Surface Science Research Centre and Department of Chemistry,
The University of Liverpool, Liverpool, L69 3BX, United Kingdom}
\affiliation{Department of Applied Physics, Chalmers University of
  Technology, SE-412 96 G\"oteborg, Sweden}
\author{Sami Paavilainen}
\affiliation{
Institute of Physics, Tampere University of Technology,
33720 Tampere, Finland}
\affiliation{Department of Applied Physics, Chalmers University of
Technology, SE-412 96 G\"oteborg, Sweden}
\author{Mats Persson}
\affiliation{
Surface Science Research Centre and Department of Chemistry,
The University of Liverpool, Liverpool, L69 3BX, United Kingdom}
\affiliation{Department of Applied Physics, Chalmers University of
  Technology, SE-412 96 G\"oteborg, Sweden}
\date{\today}
\begin{abstract}
  We have carried out a computational study of the inelastic electron
  tunneling spectrum (IETS) of the two vibrational modes of a single
  hydrogen atom on a Cu(100) surface in a scanning tunneling
  microscopy (STM) junction. This study addresses key issues about
  vibrational assignment and line shape of observed peaks in IETS
  within the framework of density functional theory calculations and
  the Lorente-Persson theory for STM-IETS. We argue that the
  observation of only a single, broad peak in the STM-IETS [L.~J.
  Lauhon and W. Ho, Phys. Rev. Lett. {\bf 85}, 4566 (2000)] is not
  caused by any symmetry restrictions or any cancellation between
  inelastic and elastic vibrational contributions for one of the two
  modes but is due to strongly overlapping superposition of the
  contributions from the two modes caused by the rather large
  instrumental broadening and the narrow vibrational energy separation
  between the modes. In particular, we find that this broadening and
  the large asymmetry of the vibrational line shapes gives rise to
  substantial apparent vibrational energy shifts of the two modes and
  decrease their apparent energy separation.
\end{abstract}
\maketitle

A most important advance in surface and nano science has been the
realization of single molecule vibrational spectroscopy and microscopy
at metal surfaces using a scanning tunneling microscope
(STM)~\cite{Stipe1998}. The unique capabilities of the STM are based
on two key ingredients: (1) the highly localized tunneling of
electrons in the space between the tip and the sample, resulting in
atomic scale resolution in imaging; (2) the non-adiabatic coupling
between tunneling electrons and the vibrations in the STM junction
giving rise to characteristic signatures in the tunneling current $I$
at biases $V$ corresponding to vibrational energies. Usually, these
thresholds are identified as peaks in the $\frac{d^2I}{dV^2}(V)$
signal. The observed vibrational energies provide unique information
about the chemical identity of single adsorbed species and can also be
used to discriminate among various isotopes. To fully exploit this new
vibrational spectroscopy, one needs physical insights from theory
about the strengths, line shapes, and lateral spatial distributions of
the vibrational signatures in the $\frac{d^2I}{dV^2}(V)$ signal.

The theory of inelastic electron tunneling spectroscopy (IETS) from
vibrations in a STM junction has been developed at various levels of
sophistication by several
groups~\cite{PerBar87,MinMak2000,Lorente2000,Mii2002}. One important
advance was made by Lorente and
Persson~\cite{Lorente2000,Lorente2000a}, who generalized the widely
used Tersoff-Hamann theory for elastic tunneling to inelastic
tunneling from vibrations and implemented the theory in density
functional theory (DFT) calculations. As demonstrated by a direct
comparison of the results from the LP theory with IETS experiments,
this theory is able to reproduce the observed strengths of various
vibrational modes for single molecules on a copper
surface~\cite{Lorente2000,Lorente2000a,Persson2004,Bocetal}. In
particular, the LP theory suggested that many modes were not observed
due to a cancellation between inelastic and elastic contributions to
the tunneling current. Furthermore, the results of the LP theory
suggested a symmetry selection or rather propensity rule for IETS
relating the symmetry of the vibrational mode, the spatial
distribution of the inelastic signal and the tip
orbital~\cite{Lorente2001}. The existence of such a rule is of key
importance in the assignment of various peaks in the IETS to
vibrational modes.

A most simple system that raises issues about the role of symmetry and
the assignment in IETS is provided by H and D atoms adsorbed on
Cu(100). In these experiments by Lauhon and Ho~\cite{Lauhon2000}, IETS
experiments were used in a beautiful manner to discriminate between
the H and D atoms, enabling the study of the isotope effect on surface
diffusion. However, they observed only a single broad peak that was
attributed to the perpendicular vibrational mode. This observation
raises questions about the origin of the large broadening of the
observed peak and why the parallel mode is not observed.

In this letter, we present a theoretical study of the IETS of H and D
atoms on a Cu(100) surface based on the LP theory and DFT
calculations. In particular, we clarify the origin of the observed
broad single peak in the IETS in terms of strongly overlapping
contributions from both the perpendicular and parallel vibrational
modes, caused by the extrinsic broadening and the intrinsic asymmetry
of the vibrational line shape. Before presenting and discussing our
results we begin by a short review of the necessary concepts and
ingredients of the theory and the calculations of IETS.

The LP theory of STM-IETS is based on the simple and physically
transparent Tersoff-Hamann theory for imaging by elastic electron
tunneling~\cite{Tersoff1985}. This theory is based on a simple tip
model, in which an emitted electron is approximated by an $s$ wave,
and the Bardeen approximation for tunneling. Using these
approximations one obtains that the differential conductance,
$\frac{dI}{dV}(V)$, at small biases, $V$, and low temperatures is
simply determined by the local density of sample states (LDOS) at the
position $\vec{r_0}$ of the tip apex as,
\begin{equation}
  \label{eq:TH}
  \frac{{\textrm{d}I}}{{\textrm{d}V}}(V)
  \propto \rho(\vec{r_0}, \epsilon_\textrm{F} +  eV),
\end{equation}
where $\epsilon_F$ is the sample Fermi energy. In an one-electron
approximation, as provided by density functional theory
$\rho(\vec{r},\epsilon)$ is given by
\begin{equation}
  \label{eq:rho}
  \rho(\vec{r}, \epsilon)=\sum_\alpha \mid\!\langle \vec{r_0}
  \!\mid\! \psi_\alpha\rangle\!\mid^2
  \delta(\epsilon_\alpha-\epsilon)
\end{equation}
where $\langle \vec{r_0} \mid \psi_\alpha\rangle$ is an one-electron
wave function with energy $\epsilon_\alpha$.

The LP theory is based on the observation that the TH theory for
elastic tunnelling can be directly generalized to inelastic tunnelling
from adsorbate vibrations by considering the many-body LDOS for the
electrons interacting with the adsorbate vibrations. This result is
based on a few physical assumptions that are fulfilled in most cases.
First, the electron-vibration coupling is assumed to be short-ranged
and limited to the sample. Second, the vibrational relaxation rate
should be much more rapid than the tunnelling rate, so that the
vibration is nearly equilibrated. Third, the electron-vibration
coupling is weak so that the vibration-induced LDOS, $\Delta\rho$ can
be evaluated by first order perturbation theory. This non-adiabatic
coupling gives rise to two distinct contributions to $\Delta\rho$ --
one from opening up an inelastic channel for tunnelling and one from
its influence on the elastic tunneling channel.

The channel for inelastic tunnelling from a vibration with vibrational
frequency $\Omega$ opens up when the bias $V$ is larger than
$\hbar\Omega/e$ and increases in general the tunnelling current. This
threshold in $I$ results in a positive peak in the line shape function
\begin{equation}\label{eq:lineshapeDef}
L_0(V) \equiv \frac{d^2I}{dV^2}(V)/\rho_\textrm{bg}
\end{equation}
where $\rho_\textrm{bg}$ is the LDOS associated with the background
differential conductance $(dI/dV)_\textrm{bg}$. The integrated
strength $\eta_{\textrm{inel}}(\vec{r}_0)$ is given by the inelastic
fraction of the tunnelling electrons with an energy larger than
$\hbar\Omega$. In the quasi-static limit, $\Omega\rightarrow 0$, this
fraction is given by the vibration-induced deformation of the vacuum
tails of the one-electron wave functions as
\begin{equation}
  \label{eq:etainel}
  \eta_{\textrm{inel}}(\vec{r_0})=\sum_\alpha \mid\!\langle
  \vec{r}_0 \!\mid\! \delta\psi_\alpha\rangle\!\mid^2
  \delta(\epsilon_\alpha-\epsilon_\textrm{F})/\rho_\textrm{bg}.
\end{equation}
Here $\langle \vec{r}_0 \mid \delta\psi_\alpha\rangle$ is the change
in $\langle \vec{r}_0 \mid \psi_\alpha\rangle$ by an rms displacement
$\sqrt{\hbar/2m\Omega}$ of the vibrational mode.

The non-adiabatic coupling of the tunnelling electrons with the
adsorbate vibration affects also the elastic tunneling channel at the
threshold for the inelastic tunneling channel through the Pauli
exclusion principle. This will show up as a decreased integrated
strength and an asymmetry of the vibrational line shape, as given by,
\begin{equation}
  \label{eq:lineshape}
  L_0(V)= \frac{\eta(\vec{r}_0)(\gamma/2)+ \chi(\vec{r}_0)
    (eV-\hbar\Omega)}{
    \pi\left((eV-\hbar\Omega)^2+(\gamma/2)^2 \right)} \ .
\end{equation}
Here the total integrated strength $\eta(\vec{r}_0)=\eta_{\rm
  inel}(\vec{r}_0)+\eta_{\rm el}(\vec{r}_0)$ has a negative
contribution from the elastic channel given by,
\begin{equation}
  \label{eq:etael}
  \eta_\textrm{el}(\vec{r}_0)= -\frac{2}
  {\rho_\textrm{bg}}\sum_\alpha \mid\!\textrm{Im}
  (\langle\delta\psi_\alpha\!\mid\! \vec{r}_0\rangle)\!\mid^2
  \delta(\epsilon_\alpha-\epsilon_\textrm{F}).
\end{equation}
The asymmetry parameter $\chi(\vec{r}_0)$ is given by,
\begin{equation}
  \label{eq:chi}
  \chi(\vec{r}_0) = -\frac{2}{\rho_\textrm{bg}} \sum_\alpha \textrm{Re}(\langle
  \vec{r}_0 \!\mid\! \delta\psi_\alpha\rangle)\textrm{Im}(\langle
  \delta\psi_\alpha \mid  \vec{r}_0\, \rangle)
  \delta(\epsilon_\alpha-\epsilon_\textrm{F})
\end{equation}
The remaining parameter determining the line shape in
Eq.~(\ref{eq:lineshape}) is the vibrational relaxation rate $\gamma$.

For high frequency adsorbate vibrations on metal surfaces such as
vibrations of H (and D) atoms on a Cu surface, $\gamma$ is dominated
by the absorption and emission of single electron-hole pairs. The
electron-hole pair contribution $\gamma_{\rm eh}$ to $\gamma$ can be
calculated also in the quasi-static limit from,~\cite{Hellsing1984}
\begin{equation}
  \label{eq:gamma}
  \gamma_\textrm{eh} = 4\pi\Omega\sum_{\alpha,\beta}
  \mid\!\langle\psi_\alpha\!\mid\! \delta v
  \mid\!\psi_\beta\rangle\!\mid^2\nonumber 
  \delta(\epsilon_\alpha-\epsilon_\textrm{F})
  \delta(\epsilon_\beta-\epsilon_\textrm{F}).
\end{equation}
Here $\delta v$ is change of the one-electron potential induced by an
rms displacement of the vibrational coordinate.

The density functional theory calculations of the electronic and
geometric structure of an H atom on a Cu(100) surface were carried out
using Vienna ab-initio simulation package {\tt
  VASP}~\cite{Kresse1996}. The electron ion core interactions were
handled by the plane wave projector augmented wave (PAW)
method~\cite{Kresse1999} and exchange-correlation effects by a
generalized gradient approximation~\cite{PW91}.  Because of the large
lateral extension of the STM image and IET signals, we have to use a
large 6x6 supercell with Cu atoms in six layers, separated by a vacuum
region corresponding to five layers of atoms.  For this large
supercell it was sufficient to sample the Brillouin zone with a 2x2x1
mesh. A cut-off energy of 20 Ry for the plane waves was found to be
sufficient. In the calculations, the H/Cu(100) system was first
geometrically optimized until forces on each ion were smaller than
0.02~eV/\AA. The technical details about the calculations of the
electron-vibration matrix elements and the vacuum tails of the wave
functions, and the handling of the discrete set of electron states can
be found in Ref.~\cite{Lorente2000a}.

The experimentally determined four-fold hollow adsorption
site~\cite{Lauhon2000} was chosen as the initial configuration in the
calculations. This site has been shown to be the energetically
preferred site also in previous DFT
studies~\cite{Sundell2004,Lai2004}. After full structural relaxation,
the vertical equilibrium distance of hydrogen from the uppermost
copper layer was 0.58~\AA. The nearest neighboring copper atoms to the
hydrogen atom relaxed 0.04~\AA\ outwards from the surface layer and
less than 0.02~\AA\ parallel the surface away from the hydrogen
adsorption site. Relaxations of the other copper atoms were found to
be much smaller. The calculated adsorption energy -0.182 eV relative
to the binding energy of the H atom in the free H$_2$ molecule
($E_{\rm ads} = E_{\rm H/Cu(100)}-E_{\rm Cu(100)}- \frac{1}{2}E_{\rm
  H_2}$) is in good agreement with the value -0.179~eV obtained by
previous DFT calculations~\cite{Sundell2004}.

\begin{figure}[t]
\includegraphics[width=8cm]{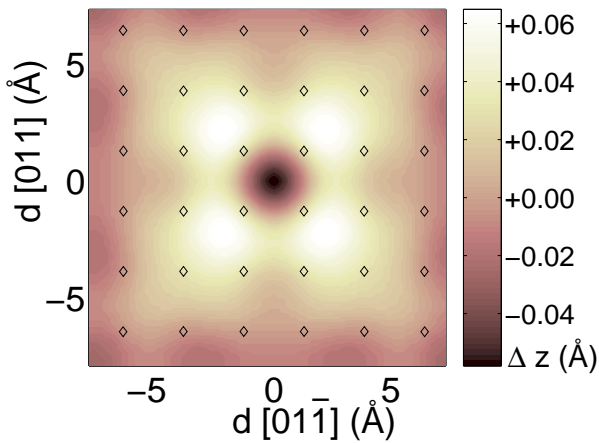}
\caption{Topographical image of constant LDOS of an H atom
  adsorbed in the hollow site on a Cu(100) surface. $\Delta z$ is the
  change in tip-surface distance from its value of 6.26~\AA\ away from
  the H atom. The positions of the Cu atoms in surface layer are
  indicated by the open diamonds.}\label{FigSTM}
\end{figure}
In Fig.~\ref{FigSTM}, we show the topographical image of constant LDOS
of the H adatom at a typical value for the tip-surface distance. The
main characteristic feature of this weakly corrugated image is the
0.05\AA\ depression at the adsorption site surrounded by four 0.06\AA\ 
protrusions centered close to the nearest neighbouring hollow sites in
the $[001]$ and $[010]$ directions. The shape of the image with the
single depression surrounded by the four protrusions is robust for all
tip-surface distances between 5 and 10\AA. However, the depression at
the hydrogen site decreases in depth with increasing tip-surface
distance $z_0$: the depth decreases from 0.07\AA\ to 0.03\AA\ when
$z_0$ increases from 5 to 8\AA. Note that the overall size of the
affected area, about 10$\times$10\AA$^2$, is very large compared to
size of hydrogen atom.  All the features in the simulated STM image
are in excellent agreement with the experimental
image~\cite{Lauhon2000}.

The origin of the protrusions in the STM image is not simply due to
the outward relaxation of the nearest neighboring copper atoms or any
H-induced state at the Fermi level. The protrusions still exist in the
simulated image of the H atom on the unrelaxed, bare Cu surface. In
accordance with the common picture of the H
chemisorption~\cite{Hje77}, the H atom induces an $s$-like state just
below the onset of the Cu $d$ band about 7 eV below the Fermi level.
The small depression in the simulated image and the surrounding
protrusions then originates from the orthogonalization of the metal
states at the Fermi level against the H-induced state and also from
the oscillatory screening response by the metal states of the H$^-$
entity.

The H adatom has two vibrational modes which in principle can
contribute to the IET vibrational spectra. In the harmonic
approximation, the calculated vibrational energies of the
perpendicular mode and the twofold degenerate parallel mode are 84 and
68 meV, respectively~\cite{enecomm}. These values are somewhat
different from the corresponding values of 76 and 70
meV~\cite{Sundell2004} (71 and 62 meV\cite{Lai2004}) found in earlier
density functional calculations which take into account anharmonic
effects. In these calculations the vibrational excitation energies
were determined from the energies of the ground state and the first
excited states of $A_1$ and $E$ symmetry character in the full
potential energy surface. These results are in better agreement
with the experimental value of 70 meV\cite{Chorkendorff1991} for the
dipole active (perpendicular) mode and will be used in the detailed
comparison with the IET experiments.

\begin{figure*}[p]
\includegraphics[width=15cm]{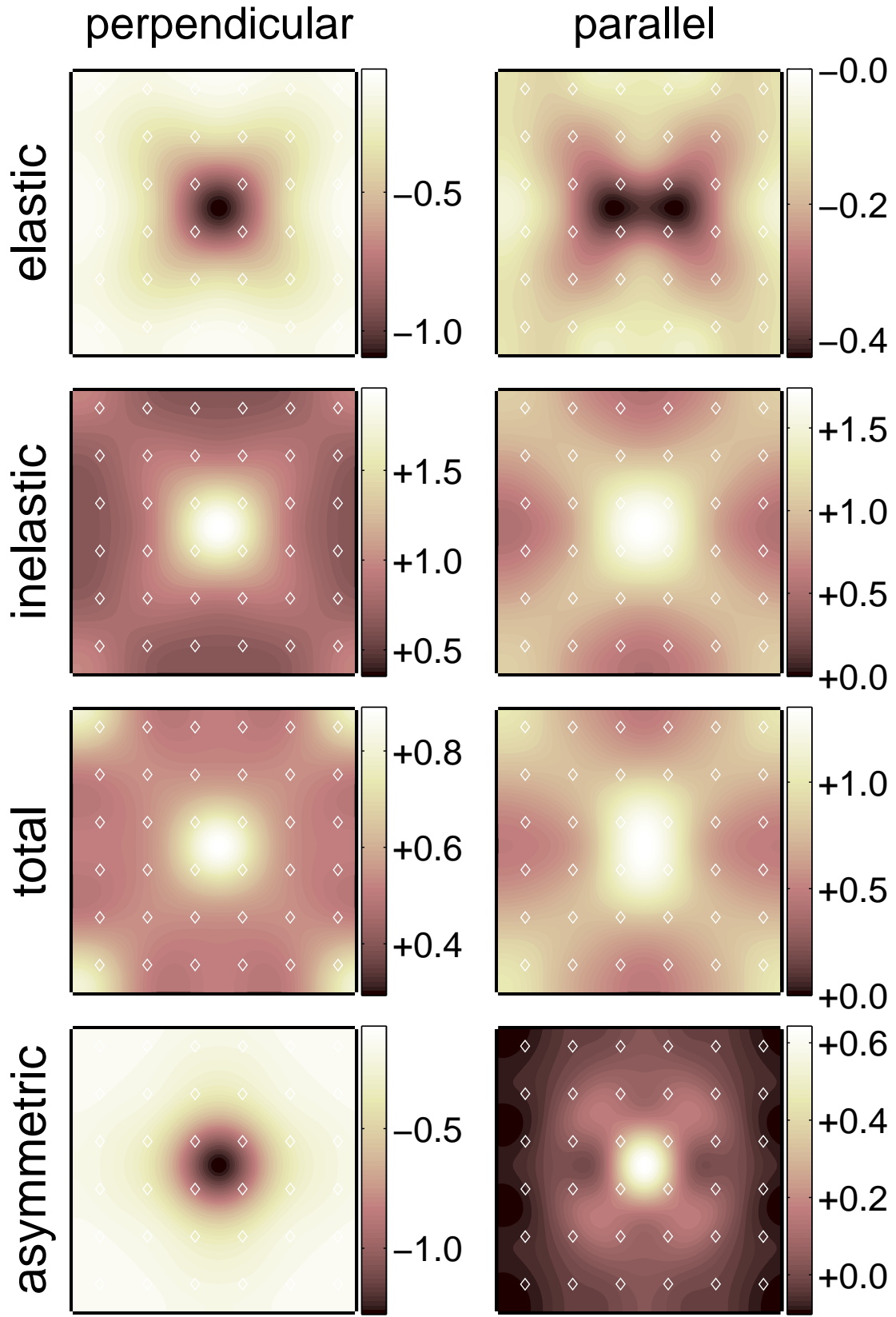}
\caption{The first three rows of the columns show the spatial maps of
  the calculated elastic, inelastic and total signal strengths for the
  perpendicular (left column) and the parallel mode vibrating in
  $[01\bar{1}]$ direction (right column), respectively. The fourth row
  of figures show the spatial map of the calculated asymmetry
  parameter for each mode. The H atom is located at the center of the
  images, and the position of the Cu atoms in the surface layer are
  indicated by open diamonds. The $x$- and $y$-axis correspond to the
  $[01\bar{1}]$ and $[011]$ crystal directions, respectively. Each
  image has the same size of 15.4\AA x15.4\AA. The tip-surface
  distance is 7 {\AA} in all figures.}
\label{Figmaps}
\end{figure*}

In Fig.~\ref{Figmaps}, we depict the two-dimensional contour plots of
the calculated IET line shape parameters
$\eta_\textrm{el}(\vec{r}_0)$, $\eta_\textrm{inel}(\vec{r}_0)$,
$\eta_\textrm{tot}(\vec{r}_0)$ and $\chi(\vec{r}_0)$, as defined in
Eqs.~\ref{eq:etainel}, \ref{eq:etael} and \ref{eq:chi}, around the H
atom for the two vibrational modes~\cite{vibcomm}. For the doubly
degenerate parallel modes it is sufficient to map out the IET
line shape parameters of the mode with a vibrational displacement in
$[01\bar{1}]$ direction since for the other degenerate mode (with a
displacement in the $[0\bar{1}1]$ direction) the parameters are simply
obtained by rotating the maps 90 degrees. The parallel mode parameters
are then obtained simply as a sum of the contributions from the two
degenerate modes.

The vibrational inelastic and elastic signals behave differently for
the two vibrational modes. In the case of the perpendicular mode, both
$\eta_\textrm{el}(\vec{r}_0)$ and $\eta_\textrm{inel}(\vec{r}_0)$ are
peaked on top of the H adsorption site, and cancel each other to a
large extent. The extreme values for $\eta_\textrm{el}(\vec{r}_0)$ and
$\eta_\textrm{inel}(\vec{r}_0)$ are -1.1\% and 2.0\%, respectively so
that $\eta_\textrm{tot}(\vec{r}_0)$ has a maximum value of 0.9\% on
top of the H atom~\cite{sigmacomm1}.  In contrast, there is no near
cancellation of $\eta_\textrm{el}(\vec{r}_0)$ and
$\eta_\textrm{inel}(\vec{r}_0)$ for the parallel mode. Thus, the
extreme value of 1.4\% for $\eta_\textrm{tot}(\vec{r}_0)$ is stronger
than for the perpendicular mode despite that the extreme values -0.4\%
and 1.4\% for $\eta_\textrm{el}(\vec{r}_0)$ and
$\eta_\textrm{inel}(\vec{r}_0)$, respectively, are both weaker than
the corresponding values for the perpendicular mode~\cite{sigmacomm1}.
Furthermore, the extreme value of $\eta_\textrm{tot}(\vec{r}_0)$ from
the twofold degenerate parallel modes will be twice as large since
this value is attained at the symmetrical site on top of the H atom.
Note that the IET line shape parameters depend somewhat on the tip-surface
distance; maps in Fig.~\ref{Figmaps} correspond to a typical distance
of 7\AA~\cite{distcomm}.

Although the perpendicular and parallel vibrational modes have
different symmetry characters ($A_1$ and $E$), the proposed symmetry
selection rule in Ref.~\cite{Lorente2001} or rather the symmetry
propensity rule for IET does not give any definite predictions about
the strengths and spatial behaviors of the inelastic and elastic
signals for these two modes in this case.  The effect of the H atom on
the states around the Fermi level is minute as demonstrated by the
small corrugation of the calculated topographical image of the H atom.
This result shows that the LDOS at the Fermi level and around the H
atom will not be dominated by states with a definite symmetry
character. Thus there will be no significant symmetry restrictions for
the IET in this case. As shown in Fig.~\ref{Figmaps}, this fact is
reflected by the spatial maps of $\eta_\textrm{tot}$ being
qualitatively same for the perpendicular and parallel vibrational
modes. Hence these maps cannot simply be used to discriminate between
these two modes.

One qualitative difference in the IET from the two vibrational modes
is the sign of the asymmetry parameter $\chi(\vec{r}_0)$, which has
important implications for the vibrational line shapes in the IETS. At
the H site, $\chi(\vec{r}_0)$ for the perpendicular mode is negative
with an extreme value of -1.2\% whereas for the parallel mode it is
positive with an extreme value of 0.7\%. Both these values are
comparable in magnitude to $\eta_\textrm{tot}(\vec{r}_0)$, and should
give rise to vibrational IETS line shapes with characteristic
asymmetries for the two modes. To show that, we need first to include
the calculated vibrational line widths, obtained from
Eq.~\ref{eq:gamma}, into Eq.~\ref{eq:lineshape} describing the
vibrational line shapes. The electron-hole pair line widths
$\gamma_{eh}$ are 1.8 and 1.1~meV for the parallel and perpendicular
modes of the H adatom, respectively~\cite{sigmacomm3}. The
corresponding relaxation rates $2.7 \cdot 10^{12} s^{-1}$ and $1.6
\cdot 10^{12} s^{-1}$ are much larger than the tunnelling rate $I/e
\sim 10^{9} s^{-1}$ corresponding to the 0.1~nA current used in the
experiments~\cite{Lauhon2000}. Thus the current-induced
non-equilibrium population of the mode should be small justifying one
of the assumptions behind the LP theory that the vibration should be
close to equilibrium.

\begin{figure}[t]
\includegraphics[width=8cm]{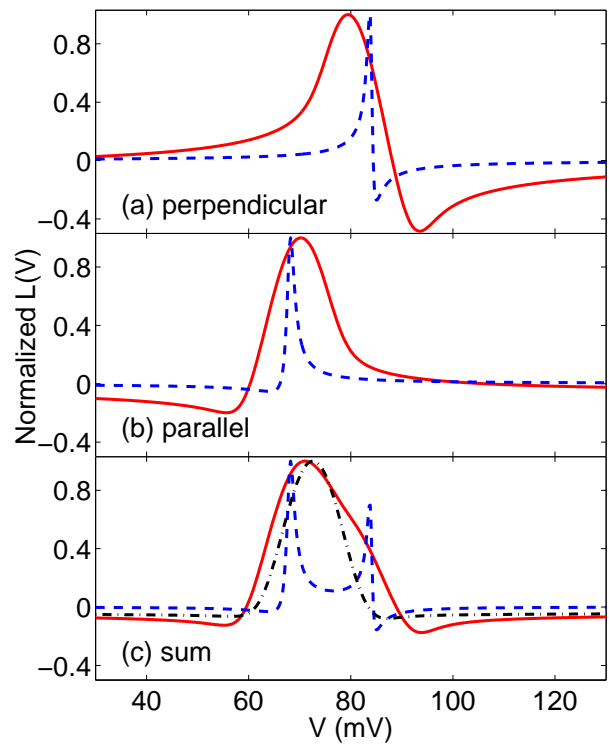}
\caption{Calculated line shapes of the normalized second order
  differential conductance for (a) the perpendicular mode, (b) the
  doubly degenerate parallel mode, and (c) the sum of the modes. The
  intrinsic line shapes are shown as dashed lines, while the line
  shapes that include temperature and modulation effects are shown as
  solid lines. All these line shapes are obtained using the calculated
  vibrational energies of 68~meV and 84~meV for perpendicular and
  parallel modes, respectively, as obtained in the harmonic
  approximation. The calculated line shape based on calculated
  vibrational energies (71~meV and 76~meV for the perpendicular and
  parallel, respectively) from Ref.~\cite{Sundell2004} that takes into
  account anharmonicity and includes temperature and modulation
  effects is shown as a dash-dotted line in (c). The conductances
  correspond to a tip-surface distance of 7 {\AA} with the tip on top
  of the H atom and $\eta=1.4 (0.9)$\%, $\chi=0.7 (-1.2)$\%, and
  $\gamma=1.79 (1.06)$~meV for the twofold degenerate parallel mode
  (perpendicular mode), respectively.} \label{lineshape}
\end{figure}

In Fig.~\ref{lineshape}, we show calculated vibrational IET line
shapes for the two modes of the H adatom with the tip on top of the H
atom. The intrinsic line shapes $L_0(V)$ (dashed lines) calculated
from Eq.~\ref{eq:lineshape} are clearly narrow and asymmetric as shown
in Figs.~\ref{lineshape}(a) and (b). Thus we would expect that the
IETS should have two clearly resolved peaks in apparent conflict with
the single broad peak in the IETS observed by Lauhon and Ho.  However,
the theory can be reconciled with experiments by taking into account
the large extrinsic broadening provided by the temperature smearing of
the tip and sample Fermi distributions, and the modulation
voltage~\cite{extbroad}. In the experiments by Lauhon and Ho, the
modulation voltage had an rms value of 7~mV and the temperature was
9~K. Including these broadenings for the two modes results in the
extrinsic line shape shown as solid lines in
Figs.~\ref{lineshape}(a)-(c). The large extrinsic broadening and the
opposite asymmetries of the two line shapes result in broad peaks that
are significantly red and blue shifted for the perpendicular and
parallel modes, respectively. In addition to decreasing the apparent
energy separation between the peaks, the broadening is so large that
the two peaks overlap strongly and can no longer be resolved as two
peaks.  This effect is even more pronounced when using the more
precise values of 71~meV and 76~meV for the vibrational excitations
obtained from Ref.~\cite{Sundell2004}.  Finally, the calculated total
strength of the IET from the two modes is about 2-3\% depending on the
distance. which is in reasonable agreement with estimate of about
1.1\% for the observed peak in the IETS. The theoretical isotope
dependence $m^{-1}$ of $\eta_{\rm tot}$ with the atomic mass $m$ is
consistent with the observed isotope dependence of the intensity for
the H and the D atom.

In summary, we have carried out a computational study of the inelastic
electron tunneling spectrum (IETS) of the two vibrational modes of a
single hydrogen atom on a Cu(100) surface in a scanning tunneling
microscopy (STM) junction. This study addresses key issues about
vibrational assignment and line shape of observed peaks in IETS and is
based on density functional theory calculations and the
Lorente-Persson theory for STM-IETS.  We argue that the observation of
only a single, broad peak is not caused by any symmetry restrictions
or any cancellation between inelastic and elastic vibrational
contributions for one of the two modes but is rather a superposition
of the contributions from the two modes caused by the large
instrumental broadening and their narrow vibrational energy
separation. In particular, we find that this broadening and the large
asymmetry of the vibrational line shapes gives rise to substantial
apparent vibrational energy shifts between the two modes and decrease
their apparent energy separation.

\section*{Acknowledgments} We acknowledge partial funding by the
EU-RTN project ``AMMIST'', the Swedish Research Council (VR) and the
Academy of Finland. Allocation of computer resources by SNAC is also
gratefully acknowledged.

%\bibliographystyle{}
%\bibliography{refsnew}

\begin{thebibliography}{99}
\bibitem{enecomm} The vibrational energies were calculated by slightly
  displacing the H atom from its adsorption site and using the
  harmonic approximation for the perturbed energies.
\bibitem{vibcomm} The calculations were carried out with the same
  displacements of the hydrogen atom as for the calculation of the
  vibrational energies.
\bibitem{sigmacomm1} The maximum values depend on the parameter
  $\sigma$ used to broaden the discrete states. For parallel mode
  using $\sigma=0.2$~eV gives a maximum $\eta_\textrm{tot}(\vec{r}_0)$
  of 1.2\% and $\sigma=0.3$~eV 0.7\%.  For perpendicular mode with
  $\sigma=0.2$~eV and 0.3~eV, the maximum of
  $\eta_\textrm{tot}(\vec{r}_0)$ are 1.9\% and 1.1\%, respectively.
  The spatial distribution of the signal is, however, not sensitive to
  the choise of $\sigma$. All the results shown correspond to
  $\sigma$=0.25eV unless mentioned otherwise.
\bibitem{distcomm}The total relative change across the perpendicular
  mode diminishes from 1.7\% to 0.9\% when increasing tip-surface
  distance from 5 \AA\ to 10 \AA. For the same distance range the
  elastic signal increases from -0.6\% to $-0.3$\%, inelastic signal
  is weakened from 2.3\% to 1.2\% and asymmetry parameter from 1.1\%
  to 0.8\%. For parallel mode, the changes are smaller; total relative
  changes goes from 1.1\% to 0.9\%, elastic signal from 1.2\% to
  0.9\%, inelastic from 2.4\% to 1.7\%, and asymmetry parameter from
  -1.5\% to -1.1\% for the corresponding distance range.
\bibitem{sigmacomm3} These values are not very sensitive to the used
  broadening of the states.  For $\sigma$=0.20(0.30)~eV the values
  were 1.85 (1.74)~meV and 0.94 (1.12)~meV, for the perpendicular and
  parallel mode, respectively.
\bibitem{extbroad} The broadenings caused by the modulation voltage
  with amplitude $\delta V$ and the tip and sample temperature $T$
  were obtained by multiplying $L_0(V)$ in the Fourier time ($\tau$)
  domain by $2J_2(\delta V \tau)/\tau^2$ and $(\pi
  k_BT\tau)^2/\sinh^2(\pi k_BT\tau)$.
\end{thebibliography}

\end{document}